\documentclass[twocolumn,preprintnumbers,nofootinbib]{revtex4-1}
\usepackage{graphicx}
\usepackage{dcolumn}
\usepackage{bm}
\usepackage{amsmath}
\usepackage{amsfonts} 
\usepackage{latexsym}
\usepackage{bbm}
\usepackage{color}
\usepackage{amssymb}
\usepackage{amsthm}
\usepackage{epsfig}

\begin{document}

\preprint{CTPU-PTC-19-15}

\title{Enhanced $Z$ boson decays as a new probe of first-order electroweak phase transition at future lepton colliders}
\author{Fa Peng Huang}
%
\author{Eibun Senaha}
\email{senaha@ibs.re.kr}

\affiliation{Center for Theoretical Physics of the Universe, Institute for Basic Science (IBS), Daejeon, 34126, Korea}

\bigskip

\date{\today}

\begin{abstract}
We study phenomenological consequences of the strong first-order electroweak phase transition 
in an extension of the standard model with an inert doublet and vector-like leptons motivated by the muon $g-2$ anomaly and dark matter.
We find that a condition for the strong first-order electroweak phase transition inevitably  
induces a large logarithmic enhancement in $Z$ boson decays,
which relegates the explanation of the anomalous muon $g-2$ at below 2$\sigma$ level.
Our analysis shows that future lepton collider experiments, especially the Giga-$Z$ at the International Linear Collider and Tera-$Z$ at the Circular Electron Positron Collider as well as Future Circular Collider have great capability to explore the nature of the electroweak phase transition, which is complementary to conventional approaches via measurements of the triple Higgs boson coupling and gravitational waves.
\end{abstract}

\maketitle

\section{Introduction}
From cosmological observations, the baryon asymmetry of the Universe (BAU) is found to be
$n_B/n_\gamma = (6.09\pm 0.06)\times 10^{-10}$~\cite{Tanabashi:2018oca}, 
where $n_B$ denotes the baryon number density and $n_\gamma$ represents the photon number density.
To obtain the observed BAU from baryon symmetric Universe, the following Sakharov's conditions~\cite{Sakharov:1967dj} must be satisfied: (i) B violation, (ii) C and CP violation and (iii) departure from thermal equilibrium.  
The last condition could be exempted if CPT is violated. 

While a plethora of baryogenesis scenarios are present in the literature~\cite{Dine:2003ax}, 
the discovery of the Higgs boson at the LHC attracts more people's attention to a scenario of electroweak baryogenesis (EWBG)~\cite{ewbg} in which the Higgs physics plays an essential role. One of the necessary ingredients for the successful EWBG is a strong first-order electroweak phase transition (SFOEWPT) that  can achieve departure from the thermal equilibrium and prevent the generated BAU from washing out.
It is shown by lattice simulations that the 125 GeV Higgs boson is too heavy to realize SFOEWPT in the standard model (SM)~\cite{sm_ewpt}, and therefore the minimal Higgs sector has to be extended by introducing, for instance, additional Higgs doublet.  Besides the baryogenesis issue, the scalar extensions of the SM are also motivated by other fundamental problems, such as the dark matter (DM), inflation and neutrino masses and mixings, and some of the  extended models could provide a solution of muon magnetic dipole moment [$(g-2)_{\mu}$] anomaly as well. 
Among such extensions, the SM augmented by an inert Higgs doublet~\cite{Barbieri:2006dq}, right-handed neutrinos and vector-like leptons (denoted as VLIDM in short) has been studied from the viewpoints of DM and neutrino physics~\cite{Borah:2017dqx}, 
or DM and $(g-2)_{\mu}$ anomaly~\cite{Calibbi:2018rzv}, or DM and inflation~\cite{Borah:2018rca}.
One can expect that SFOEWPT would still be possible in the VLIDM as is the ordinary inert Higgs doublet model.
However, so far there is no explicit demonstration and compatibility of SFOEWPT with other observables, especially $(g-2)_{\mu}$ is not clear.
 
In this paper, we study SFOEWPT and its compatibility with the $(g-2)_{\mu}$ explanation
in the VLIDM, and also discuss phenomenological consequences,
focusing particularly on the correlation between SFOEWPT and the $Z$ boson decays. 
Here, we consider a case in which the vector-like leptons preferentially couple to muons,
which is motivated by the $(g-2)_\mu$ explanation. 

We point out that a condition for SFOEWPT inevitably leads to sizable radiative corrections to $Z\to \mu^+\mu^-$ due to
a logarithmic enhancement factor, whereas $(g-2)_\mu$ , by contrast, is suppressed,
preventing one from explaining the $(g-2)_\mu$ anomaly within 2$\sigma$ level.
Since the essential point in this correlation is a mass splitting between the neutral scalars in the same multiplet, our findings would hold in other models as long as the mass splitting is crucial for realizing SFOEWPT and EWBG

We also show that the regions of SFOEWPT and DM in our scenario can be thoroughly probed by the future lepton collider experiments, especially the precise measurements of the $Z$ boson, such as a Giga-$Z$ option at the International Linear Collider (ILC)~\cite{Irles:2019xny}
as well as Tera-$Z$ phase at the Circular Electron-Positron Collider (CEPC)~\cite{CEPCStudyGroup:2018ghi}
and Future Circular Collider (FCC-ee)~\cite{Blondel:2018mad},
which plan to produce around $10^{9}$ and $10^{12}$ $Z$ bosons, respectively.
\footnote{For earlier studies on DM and/or $(g-2)_\mu$-favored regions at the $Z$ factories, see, {\it e.g.}, Refs.~\cite{Cao:2016qgc, ColuccioLeskow:2016dox,Kowalska:2017iqv,Liu:2017zdh}.}
This would be a new  approach to explore the nature of EWPT 
along with conventional probes by the triple Higgs boson coupling~\cite{Grojean:2004xa,Kanemura:2004ch} and gravitational waves~\cite{Huang:2017rzf}.

The paper is organized as follows. In Sec.~\ref{sec:model}, we describe the VLIDM and give a brief overview of DM physics in this model.
In Sec.~\ref{sec:Pheno_EWPT}, we derive the condition of SFOEWPT and show its impacts on
the $Z$ decays and $\delta a_\mu$ analytically. The quantitative studies of the correlations are presented in Sec.~\ref{sec:numerical_analysis}, and 
Sec.~\ref{sec:conclusion} is devoted to conclusion and discussions.
One-loop functions appearing in the $Z$ boson decays are listed in Appendix \ref{LF}.

\section{The model}\label{sec:model}
We study the model in which the inert Higgs doublet ($\eta$), right-handed neutrinos ($N_{Ri=1-3}$)~\cite{Ma:2006km} and vector-like leptons ($E_{i=1-3}$) are added to the SM~\cite{Borah:2017dqx,Calibbi:2018rzv,Barman:2018jhz}.
The SM quantum numbers and the $Z_2$ parity of each field are assigned as 
 $(\boldsymbol{1}, \boldsymbol{2}, 1/2, -)$ for $\eta$, $(\boldsymbol{1}, \boldsymbol{1}, 0, -)$ for $N_{Ri}$
 and $(\boldsymbol{1}, \boldsymbol{1}, -1, -)$ for $E_i$, respectively.
This model offers the possible explanation for the DM, neutrino masses/mixings and the $(g-2)_\mu$ anomaly.
\footnote{In this model, the neutrino masses are generated by the same mechanism as the Ma model~\cite{Ma:2006km}, and the vector-like leptons do not play a significant role.}
Moreover, if we further assume a non-minimal coupling of $\eta$ to gravity, 
a successful scenario of inflation can also be accommodated~\cite{Borah:2018rca}. 
Owing to the $Z_2$ parity, the lightest $Z_2$-odd particle can be stable and becomes the DM candidate. 
In this work, we focus on the case of the scalar DM by assuming the right-handed neutrinos are heavy enough and thus omitted hereafter.

The new vector-like lepton interactions can be written as
\begin{equation}
-{\cal L} \supset  m_{E_i} \bar{E}_{iL} E_{iR} + y_{ij} \, \bar{\ell}_{iL} \eta E_{jR}  + \text{h.c.}  \ ,
\end{equation}
where $y_{ij}$ are the general 3-by-3 complex matrices that may
provide the necessary CP-violating sources for EWBG~\cite{bar}.\footnote{In principle, there is a possibility of leptogenesis~\cite{Fukugita:1986hr} (for recent studies, see, {\it e.g.}, Refs.~\cite{Borah:2018rca,Hugle:2018qbw}). The amount of BAU strongly depends on parameters in the lepton sector. In our scenario, the BAU is not generated via leptogenesis by assumption.} 
Since we focus on the $(g-2)_\mu$-favored region, we assume that $y_{\mu E_i}\neq0$ and other elements are negligibly small. Moreover, we set $y_{\mu E_1}=y_{\mu E_2}=y_{\mu E_3}\equiv y_{\mu E}$ and $m_{E_1}=m_{E_2}=m_{E_3}\equiv m_{E}$ for the sake of simplicity. 

The tree-level scalar potential is given by
\begin{equation}
\begin{aligned}
V_0(\Phi,\eta)&= \mu_1^2|\Phi|^2 +\mu_2^2|\eta|^2+\frac{\lambda_1}{2}|\Phi|^4+\frac{\lambda_2}{2}|\eta|^4+\lambda_3|\Phi|^2|\eta|^2 \nonumber\\
&\quad+\lambda_4|\Phi^\dag \eta|^2 + \left[\frac{\lambda_5}{2}(\Phi^\dag \eta)^2 + \text{h.c.}\right],
\end{aligned}
\label {c}
\end{equation}
where $\Phi$ is the SM Higgs doublet that develops a vacuum expectation value (VEV) with $v=246$ GeV.
The masses of the physical scalar particles  at tree level can be written as
\begin{align}
m_h^2 &= \lambda_1 v^2 ,\nonumber\\
m_{H^{\pm}}^2 &= \mu_2^2 + \frac{1}{2}\lambda_3 v^2 , \nonumber\\
m_{H}^2 &= \mu_2^2 + \frac{1}{2}(\lambda_3+\lambda_4+\lambda_5)v^2=m^2_{H^\pm}+
\frac{1}{2}\left(\lambda_4+\lambda_5\right)v^2  , \nonumber\\
m_{A}^2 &= \mu_2^2 + \frac{1}{2}(\lambda_3+\lambda_4-\lambda_5)v^2=m^2_{H^\pm}+
\frac{1}{2}\left(\lambda_4-\lambda_5\right)v^2,
\label{ms}
\end{align}
where $m_h$ is the SM Higgs boson mass, $m_H$ and $m_A$ are the masses of the CP-even and -odd scalar particles from the inert Higgs doublet $\eta$, respectively and $m_{H^{\pm}}$ is the charged Higgs boson mass.
Without loss of generality, we consider that $\lambda_4+\lambda_5 <0$ and $\lambda_5 <0$,
which renders the CP-even scalar $H$ the lightest $Z_2$-odd particle and hence the stable DM candidate.
As is known, the contributions from the extra scalars to the $T$ parameter becomes zero if $m_{H^\pm}=m_H$ or $m_{H^\pm}=m_A$~\cite{Barbieri:2006dq}.
Although the small deviations from those mass relations are still experimentally allowed, we choose $m_{H^{\pm}}=m_A$, corresponding to $\lambda_4=\lambda_5<0$, in order to make our discussion simpler.
We also note that in parameter space of our interest, $S$ parameter~\cite{Barbieri:2006dq} is small.

From the Planck 2018 data~\cite{Aghanim:2018eyx},
the observed DM abundance is determined as 
\begin{equation}
\Omega_{\text{DM}} h^2 =  0.11933 \pm 0.00091.
\label{dm_relic}
\end{equation}
If the DM mass is less than $W$ boson mass, the main DM annihilation processes are the $h$-mediated $s$ channel and the vector-like lepton-mediated $t$ channel.
Its cross section is approximately given by
\begin{align}
\sigma v_{\text{rel}}
& \simeq \frac{N_C^f}{16\pi}\frac{\lambda_L^2m_f^2\beta_f^3}{(s-m_h^2)^2+m_h^2\Gamma_h^2} +\frac{|y_{\mu E}|^4}{60\pi m_H^2}\frac{v_{\text{rel}}^4}{(1+r_{E})^4},\label{DM-anni}
\end{align}
where $v_{\text{rel}}$ denotes the relative velocity of the DM, $f$ denote the SM fermions, $N_C^f$ are the color degrees of freedom of $f$, $\lambda_L=\lambda_3+\lambda_4+\lambda_5$, $\Gamma_h$ is the width of the Higgs boson, $r_{E}=m_{E}^2/m_H^2$, $s=4m_H^2/(1-v_{\text{rel}}^2/4)$, $\beta_f  = \sqrt{1-4m_f^2/s}$.
Although the $t$-channel process is $d$-wave suppressed, it would come into play if $|y_{\mu E}|\gtrsim 1$~\cite{Calibbi:2018rzv}.

As for the DM direct detection, the recent XENON1T data put a strong constraint on the DM-nucleon spin-independent elastic scatter cross-section $\sigma_{\text{SI}}$~\cite{Aprile:2018dbl}.
For instance, the most excluded region at $90\%$ confidence level reaches $\sigma_{\text{SI}}=4.1 \times 10^{-47}~\text{cm}^2$ with the DM mass of 30 GeV.
Therefore, for a light DM, the direct detection data favor the so-called Higgs funnel region where the DM mass is close to half of the Higgs mass, 
namely, $m_H  \simeq m_h/2\simeq 63$ GeV.
In this model, the cross-section $\sigma_{\text{SI}}$ is approximated as
\begin{eqnarray}
	\sigma_{\rm SI} \simeq  \frac{\lambda_L^2f_N^2}{4\pi} \left(\frac{m_N^2}{m_{H} m_h^2}\right)^2,
	\label{sigmaSI}
\end{eqnarray}
where $f_N \simeq 0.3$.
To evade the current DM direct detection constraints in this Higgs funnel region, $\lambda_L \lesssim 0.003$ is required,
which implies that 
\begin{align}
m_H^2 \simeq \mu_2^2,\quad m_A^2 \simeq m_H^2+\frac{\lambda_3}{2}v^2.
\label{mass_spectrum}
\end{align}
As discussed below, $\lambda_3$ has to be $\mathcal{O}(1)$ in magnitude to achieve SFOEWPT~\cite{Chowdhury:2011ga,Borah:2012pu,Gil:2012ya,Cline:2013bln,AbdusSalam:2013eya,Blinov:2015vma,Cao:2017oez,Laine:2017hdk,Senaha:2018xek,Kainulainen:2019kyp}.

Although Eqs.~(\ref{DM-anni}) and (\ref{sigmaSI}) make it easy to see the model parameter dependences, 
we use micrOMEGAs~\cite{Barducci:2016pcb}  in order to get more precise values of 
$\Omega_{\text{DM}}h^2$ and $\sigma_{\text{SI}}$.
We have confirmed that our numerical results are consistent with those in Ref.~\cite{Calibbi:2018rzv}
when we take their input parameters.

\section{Strong first-order phase transition and its implications for $Z$ boson decays and $(g-2)_\mu$}\label{sec:Pheno_EWPT}
In EWBG, the BAU arises via $B$-violating processes (sphaleron processes) in the symmetric phase. 
To maintain the generated BAU in the broken phase, the sphaleron processes must be sufficiently suppressed.
This is realized if the sphaleron energy, which is proportional to the Higgs VEV at finite temperature, becomes
large enough. Conventionally, the condition of SFOEWPT is approximately described by 
\begin{align}
v_C/T_C\gtrsim 1,\label{bnpc}
\end{align}
where $T_C$ denotes the critical temperature at which two degenerate minima coexist in the finite-temperature effective scalar potential and $v_C$ is the corresponding VEV in the broken phase. 

Let us consider how the condition of (\ref{bnpc}) is satisfied in this model. 
The effective potential with high-temperature expansion takes the form
\begin{align}
V_{\text{eff}}(\varphi; T_C) = \frac{\lambda_{T_C}}{4}\varphi^2(\varphi-v_C)^2,
\quad v_C=\frac{2ET_C}{\lambda_{T_C}},\label{1LHTE}
\end{align}
where $\varphi$ is the classical background field of the SM-like Higgs, $E$ denotes the coefficient of the $\varphi^3$ term and $\lambda_{T_C}$ is the quartic coupling at $T_C$, 
which is more or less fixed by the Higgs boson mass. Therefore, $E$ has to be enhanced in order to satisfy Eq.~(\ref{bnpc}). As is well known, the bosonic particles can contribute to $E$.
In this model, $E$ would be enhanced if $\mu_2^2\ll \lambda_3v^2/2$, which enforces the large mass splitting
among the neutral scalars of the inert doublet, {\it i.e.},
\begin{align}
m_H \ll m_A. \label{mHllmA}
\end{align}
To what extent the mass splitting is needed depends on the condition (\ref{bnpc}).
Even though Eq.~(\ref{1LHTE}) is useful to understand the essence of SFOEWPT, 
the high-temperature expansion is not always valid. We therefore numerically evaluate $v_C/T_C$ in Sec.~\ref{sec:numerical_analysis}. Note that since the vector-like leptons do not generate the thermal cubic term, SFOEWPT in the parameter space of our interest is essentially the same as that in the ordinary inert doublet model.
Note also that too large mass splitting breaks the perturbativity of $\lambda_3$ as seen from Eq.~(\ref{mass_spectrum}).
In what follows, we discuss the implications of the condition (\ref{mHllmA}) for the $Z$ boson decays and $(g-2)_\mu$.

\begin{figure}[t]
\begin{center}
\includegraphics[scale=0.33]{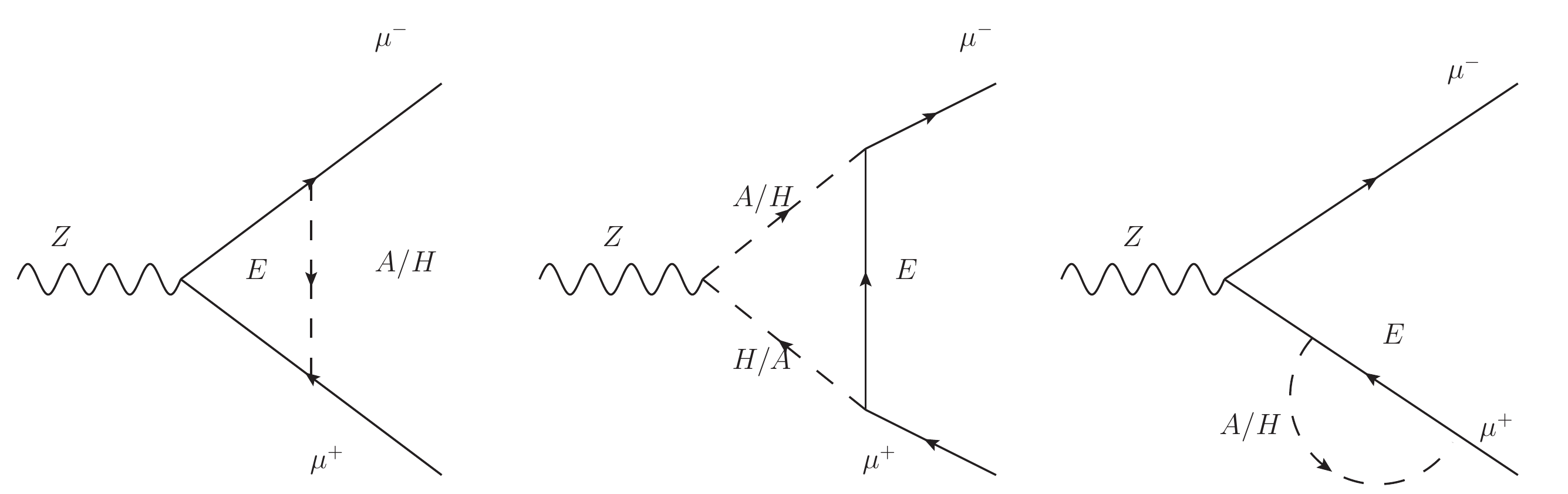}
\end{center} 
\caption{Representative Feynman diagrams  for the one-loop corrections to $Z\to  \mu^+\mu^-$. }
\label{zmmloop}
\end{figure}
Let us parametrize the $Z$ boson couplings to fermions as
\begin{align}
\mathcal{L}& = -g_ZZ_\mu \bar{f}\gamma^\mu\Big[g_{Z\bar{f}f}^LP_L+g_{Z\bar{f}f}^RP_R \Big]f,
\end{align}
where $g_Z = g_2/c_W$ with $g_2$ being the SU(2) gauge coupling and $c_W$ the cosine of the weak mixing angle. With those $Z$ boson couplings, the partial decay width of $Z\to \ell^+\ell^-$ can be written as
\begin{align}
\Gamma(Z\to \ell^+\ell^-) &= \frac{m_Z}{24\pi}g_Z^2\Big[|g_{Z\bar{\ell}\ell}^L|^2+|g_{Z\bar{\ell}\ell}^R|^2 \Big],
\end{align}
where the SM lepton masses are ignored. 
We parametrize the new physics effects as $g_{Z\bar{\ell}\ell}^{L,R} = g_{Z\bar{\ell}\ell}^{L,R,\text{SM}}+\Delta g_{Z\bar{\ell}\ell}^{L,R}$. It is straightforward  to calculate the new physics effects by evaluating the Feynman diagrams
as depicted in Fig.~\ref{zmmloop} (the leg correction to $\mu^-$ also exists.).
In the current model, $\Delta g_{Z\bar{\mu}\mu}^{R}=0$ and
\begin{align}
\Delta g_{Z\bar{\mu}\mu}^{L}
&= \frac{3|y_{\mu E}|^2}{32\pi^2} 
\Bigg[\tilde{F}_3(m_{E},m_H,m_A) \nonumber\\
&\hspace{1cm}
	+\sum_{\phi=H,A}\bigg\{
	\left(-\frac{1}{2}+s_W^2\right)F_2(m_{E}, m_\phi) \nonumber\\
&\hspace{4cm}
	+s_W^2F_3(m_{E}, m_\phi)	
	\bigg\} \Bigg],
\end{align}
where the loop functions $F_2$, $F_3$ and $\tilde{F}_3$ are listed in Appendix~\ref{LF}.
One can show that $\Delta g_{Z\bar{\mu}\mu}^{L}\to 0$ for $m_\phi(\simeq m_Z) \ll m_E$, $m_E(\simeq m_Z)\ll m_\phi$, and $m_Z\ll m_\phi=m_E$ as long as  $m_\phi=m_H=m_A$. 
For $m_H, m_E \ll m_A$, however, $\Delta g_{Z\bar{\mu}\mu}^{L}$ would be enhanced by $\ln(m_A^2/m_H^2)$
that arises from the correction of the middle triangle  diagram and the right leg corrections of $\mu^\pm$ in Fig.~\ref{zmmloop}. 
As a result, the corrections to $\Gamma(Z\to \mu^+\mu^-)$ in this limit is cast into the form
\begin{align}
\lefteqn{\Delta \Gamma(Z\to \mu^+\mu^-)} \nonumber\\
&\simeq \frac{m_Z^2g_Z^2|y_{\mu E}|^2}{128\pi^3}
\left(-\frac{1}{2}+s_W^2\right) \left[C+\frac{1}{4}\ln\frac{m_A^2}{m_H^2}\right],
\end{align}
where $C$ denotes non-logarithmic contributions.
We are aware of that such a non-decoupling behavior by the mass splitting is already noticed
in the calculation of $Z\to b\bar{b}$ in the two-Higgs doublet model~\cite{Haber:1999zh},
and more recently in the study of $(g-2)_\mu$ in the lepton-specific two-Higgs doublet model~\cite{Chun:2016hzs}.
\footnote{In the lepton-specific model, leptonic $\tau$ decays are also enhanced by the similar logarithmic contribution, which is as important as the $Z$ boson decays~\cite{Chun:2016hzs,Abe:2015oca,Wang:2018hnw}.}
Nevertheless, to our best knowledge, the importance of its correlation with SFOEWPT 
has not been well-recognized in the literature and therefore
detailed numerical studies will be conducted below.

Since the vector-like leptons couple only to $\mu^\pm$ in this model,  
the lepton flavor universality of $Z$ boson decays is violated. 
We thus utilize 
\begin{align}
R_{\mu/e} = \frac{\Gamma(Z\to \mu^+\mu^-)}{\Gamma(Z\to e^+e^-)}
\end{align}
to test this model precisely. Its current experimental value is $R_{\mu/e}^{\text{EXP}} = 1.0009 \pm 0.0028$~\cite{Tanabashi:2018oca}.
Let us define the deviation of $R_{\mu/e}$ from the SM value as
\begin{align}
\Delta R_{\mu/e}
&\equiv \frac{R_{\mu/e}-R_{\mu/e}^{\text{SM}}}{R_{\mu/e}^{\text{SM}}} \nonumber\\
& \simeq \frac{2g_{Z\bar{\mu}\mu}^{L,\text{SM}}\text{Re}\big(\Delta g_{Z\bar{\mu}\mu}^{L}\big)
	+|\Delta g_{Z\bar{\mu}\mu}^{L}|^2}{|g_{Z\bar{\mu}\mu}^{L,\text{SM}}|^2+|g_{Z\bar{\mu}\mu}^{R,\text{SM}}|^2},
\end{align}
where $g_{Z\bar{\mu}\mu}^{L,\text{SM}}\simeq -0.27$ and $g_{Z\bar{\mu}\mu}^{R,\text{SM}}\simeq0.23$.
As the experimental constraints, we require that $\Delta R_{\mu/e}<2.8\times 10^{-3}$.

As is the case of $\Gamma(Z\to \mu^+\mu^-)$, $\Gamma(Z\to \nu\bar{\nu})$ is also modified by the new particles as
\begin{align}
\lefteqn{\Delta g_{Z\bar{\nu}\nu}^L} \nonumber\\ 
&= \frac{3|y_{\mu E}|^2}{16\pi^2}
\bigg[
	s_W^2\left\{F_3(m_{E},m_{H^\pm})+\tilde{F}_3(m_{E},m_{H^\pm},m_{H^\pm}) \right\}
	 \nonumber\\
&\hspace{1.3cm}
	+\frac{1}{2}\left\{F_2(m_{E}, m_{H^\pm})-\tilde{F}_3(m_{E},m_{H^\pm},m_{H^\pm}) \right\}
\bigg], 
\end{align}
and $\Delta g_{Z\bar{\nu}\nu}^R = 0$. 
Unlike the $Z\to \mu^+\mu^-$ case, $\Delta g_{Z\bar{\nu}\nu}^L$ does not have the logarithmic enhancement due to the absence of the mass splitting. 
Since this quantity is always numerically unimportant, we do not discuss it henceforward.

Here, we also make a comment on other experimental constraint, especially the $W$-$\mu$-$\nu_\mu$ coupling ($\Delta g_{W\mu\nu}^L$) whose deviation from the SM induces the lepton flavor non-universality in the muon decay.
As with $\Delta g_{Z\bar{\mu}\mu}^L$, $\Delta g_{W\mu\nu}^L$ would receive the logarithmic enhancement 
of $\ln(m_A^2/m_H^2)$ in the limit of $m_H\ll m_A$. In our parameter space, however, $R_{\mu/e}$ gives
the stronger bound than this constraint so that we do not discuss it in the following.

\begin{figure}[t]
\begin{center}
\includegraphics[scale=0.35]{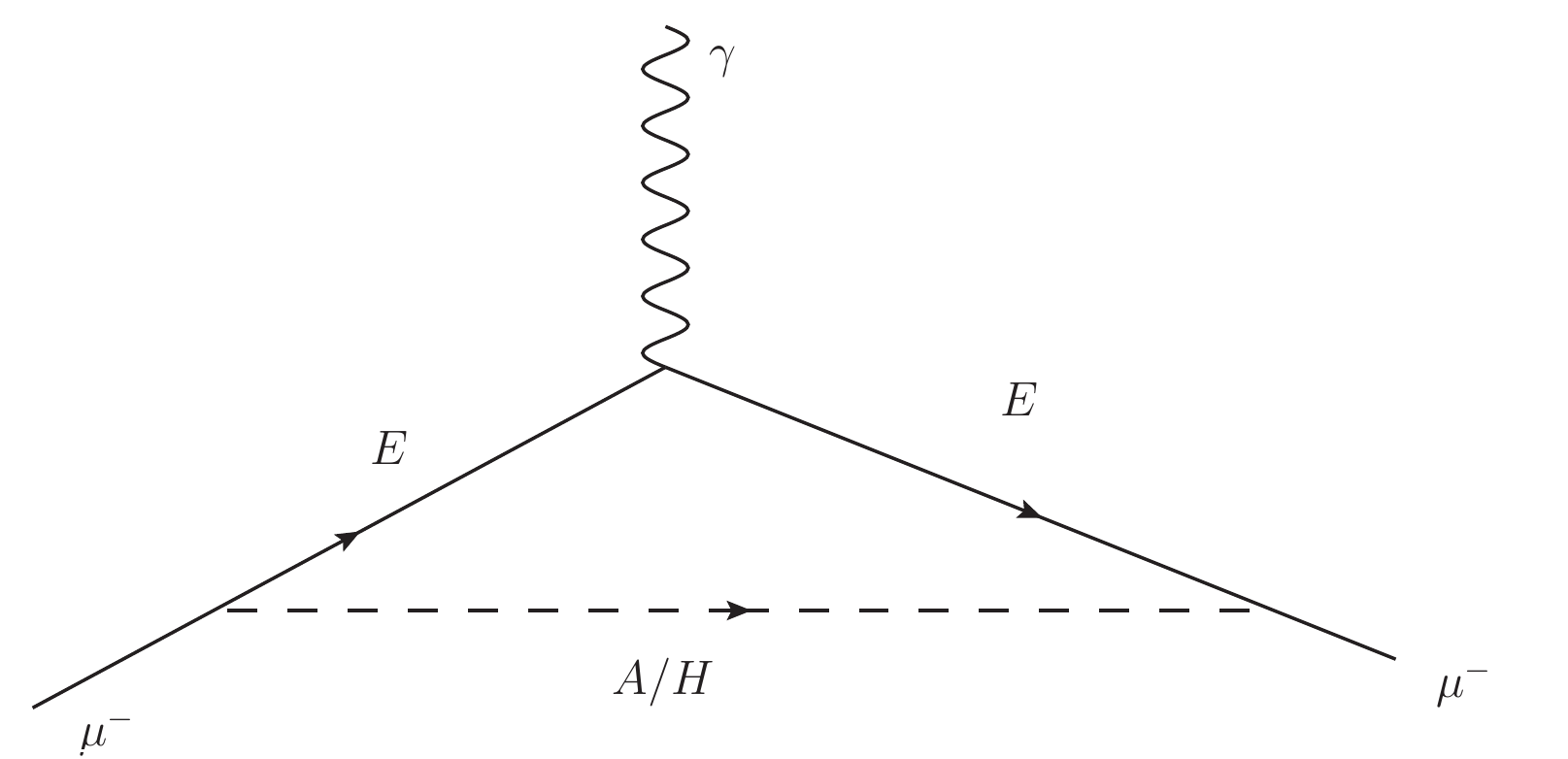}
\end{center} 
\caption{Schematic  Feynman diagrams for the one-loop contributions to $(g-2)_\mu$.}
\label{g_2}
\end{figure}
Now we move to consider the new physics effects on $(g-2)_\mu$ in the case of $m_H\ll m_A$.

The discrepancy of $(g-2)_\mu$ between the experimental value and the SM prediction 
is estimated as~\cite{Hagiwara:2011af}
\begin{align}
\delta a_\mu = a_\mu^{\rm EXP}-a_\mu^{\rm SM} = (26.1\pm 8.0) \times 10^{-10}.
\end{align}
In this model, the one-loop contributions to $(g-2)_\mu$ arise from the diagram shown in Fig.~\ref{g_2},
from which it is found to be~\cite{Leveille:1977rc}
\begin{align}
\delta a_\mu = \sum_{\phi=H,A}\frac{3|y_{\mu E}|^2}{32\pi^2}S_1(r_{E\mu}, r_{\phi\mu}),
\end{align}
where $r_{E\mu}=m_{E}^2/m_{\mu}^2$, $r_{\phi \mu}=m_{\phi}^2/m_{\mu}^2$ with $m_\mu$ being the muon mass
and
\begin{align}
S_1(r_1, r_2) = \int_0^1 dx~\frac{x^2(1-x)}{x(x-1)+xr_1+(1-x)r_2}.
\end{align}
For $m_{E}\simeq m_H \ll m_A$, $\delta a_\mu$ is approximated as
\begin{align}
\lefteqn{\delta a_\mu}\nonumber\\
&\simeq \frac{3|y_{\mu E}|^2}{32\pi^2}
\left[
\frac{m_\mu^2}{12m_H^2}+\frac{m_\mu^2}{m_A^2}\left\{\frac{1}{3}+\frac{m_E^2}{m_A^2}\left(\frac{11}{6}+\ln\frac{m_E^2}{m_A^2}\right)\right\}
\right].
\end{align}
Therefore, the heavy particle simply decouples, which is in stark contrast to the case of $Z\to \mu^+\mu^-$.
This is understandable since the logarithmic enhancement is originated from 
the vertex correction of the middle diagram and the leg corrections of $\mu^\pm$ in Fig.~\ref{zmmloop} 
while their counterparts are absent in the diagram of $\delta a_\mu$.

\section{Numerical analysis}\label{sec:numerical_analysis}
Before showing our numerical results, we outline the current experimental constraints.
As mentioned above, the DM data enforce that $m_H\simeq m_h/2$ and $\lambda_L\lesssim 0.003$ for the light DM case.
As for the LHC constraints, there are two important processes (for recent studies, see, {\it e.g.}, Refs.~\cite{Belyaev:2016lok,Bhardwaj:2019mts}). 
One is the dimuon plus missing energy (MET) process, $q \bar{q}\to E^+E^- \to \mu^+ \mu^-+\text{MET}$,
from which the vector-like lepton mass is bounded as $105~\text{GeV}\lesssim m_E \lesssim 125~\text{GeV}$~\cite{Calibbi:2018rzv}. The other is the monojet plus MET, $gg \to g h \to g HH$.
However, this cross section would be suppressed by $\lambda_L^2$ if the aforementioned DM constraint is taken into account.  
Other than those, the potentially relevant constraint is the signal strength of the Higgs boson decays to two photons ($\mu_{\gamma\gamma}$), which can be affected by the charged Higgs bosons.
In parameter space of SFOEWPT, it is known that $\mu_{\gamma\gamma}\simeq 0.9$~\cite{Blinov:2015vma}, which is consistent with the current LHC data $\mu_{\gamma\gamma}^{\text{ATLAS}}=0.99_{-0.14}^{+0.15}$~\cite{Aaboud:2018xdt} and $\mu_{\gamma\gamma}^{\text{CMS}}=1.18_{-0.14}^{+0.17}$~\cite{Sirunyan:2018ouh} within 2$\sigma$ level.

As a benchmark scenario, we consider
\begin{align}
m_E=(105 -125)~\text{GeV},\quad |y_{\mu E}|=1.0, 0.5~\mbox{and}~0.3.
\end{align}
The DM relic abundance is always satisfied by judiciously choosing $m_H$ and $\lambda_L$.
For instance, for $m_E=110$ GeV and $|y_{\mu E}|=0.5$, the choice of $m_H=62.55$~GeV and $\lambda_L=0.001$ gives $\Omega_{\text{DM}}h^2=0.12$ and 
$\sigma_{\text{SI}}=8.7 \times 10^{-48}~\text{cm}^2$. Here, we set $m_A=m_{H^\pm}=300$ GeV and $\lambda_2=0.3$,
though they are not sensitive to the results.

It would be nice if the above scenario can be tested at the LHC Run-3 or high luminosity (HL)-LHC. 
However, it might be difficult since such light vector-like leptons could escape from the searches via 
the soft lepton plus MET as well as the dilepton plus MET~\cite{Calibbi:2018rzv}. 
Furthermore, the monojet plus MET process can constrain $\lambda_L$ only down to $\mathcal{O}(0.1)$ at the HL-LHC~\cite{Belyaev:2016lok}, which is still way above the range of our interest. 
It is definitely worth conducting dedicated studies taking all the detailed information into account~\cite{Barman:2018jhz,Bhardwaj:2019mts}. 
In this work, however, we consider detectability at future lepton colliders instead,
which can offer more robust tests of the scenario.

We firstly consider the case in which the vector-like leptons cannot be pair produced 
at the future lepton collider with the center of mass energy of 240 GeV (CEPC, FCC-ee) or 250 GeV (ILC), respectively, {\it i.e.}, $m_E=120$ GeV or 125 GeV.
Since physics discussion would not differ between the two cases, we only present the $m_E=120$ GeV case below. 
After that, we also consider the case in which the vector-like leptons can be directly produced at those colliders.

\begin{figure}[t]
\begin{center}
\includegraphics[scale=0.65]{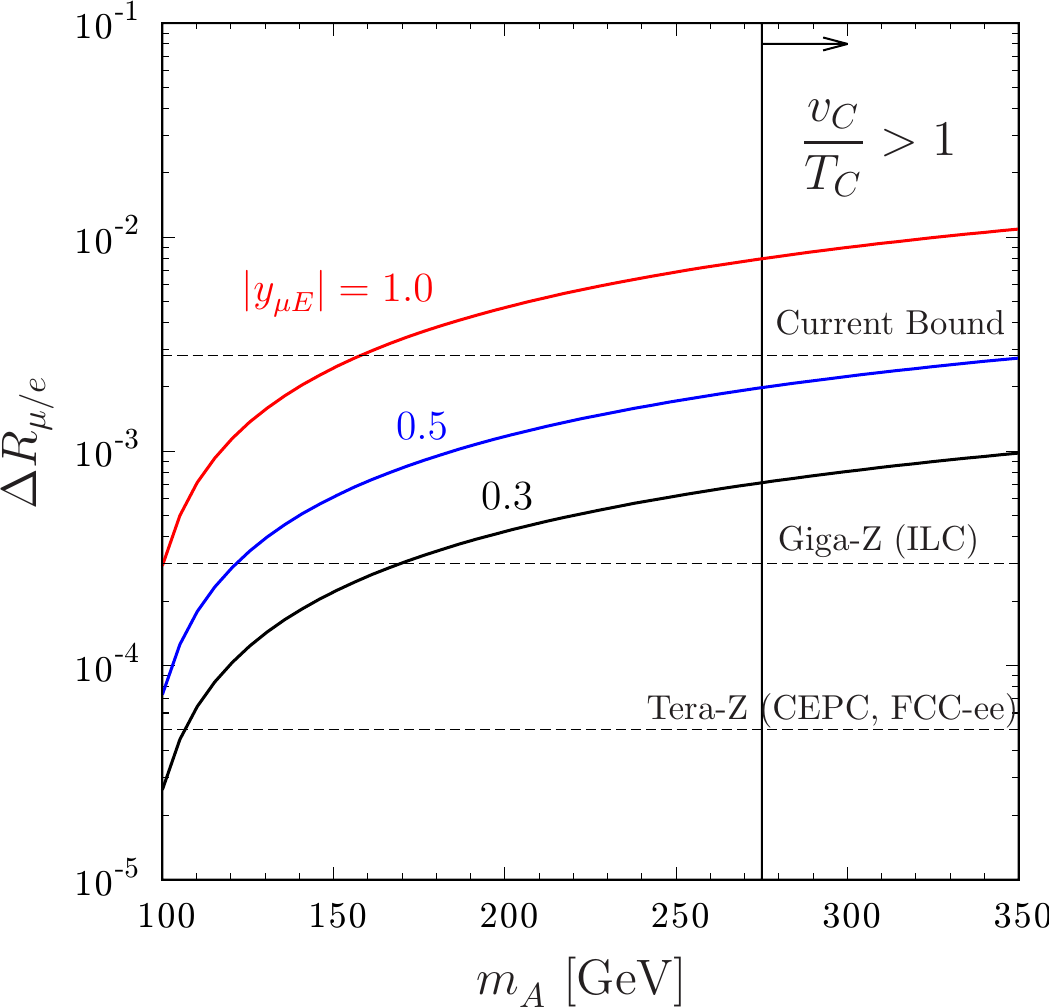} 
\end{center} 
\caption{Deviation of the lepton flavor universality in the $Z$ boson decays $\Delta R_{\mu/e}$ as a function of $m_A$ with $m_E=120$ GeV.
The solid curves in red, blue and black correspond to the deviation $\Delta R_{\mu/e}$  in the $Z$ boson decay for $|y_{\mu E}|=1.0$, 0.5 and 0.3, respectively.
The upper, middle and lower horizontal dashed lines represent the current bound, future Giga-$Z$ sensitivity and Tera-$Z$ sensitivity of $\Delta R_{\mu/e}$, respectively.
The region of SFOEWPT is the right-hand side of the vertical line.  The greater $v_C/T_C$ corresponds to the larger enhancement of $Z\to \mu^+\mu^-$.}
\label{Rmue_MA}
\end{figure}

In Fig.~\ref{Rmue_MA}, $\Delta R_{\mu/e}$ is plotted as a function of the heavy scalar mass $m_A$.
The current upper bound of $\Delta R_{\mu/e}$ is represented by the upper horizontal dashed line.
The solid curves in red, blue and black correspond to the deviation of $\Delta R_{\mu/e}$ in the $Z$ boson decay for $|y_{\mu E}|=1.0$, 0.5 and 0.3, respectively.
In any cases, $\Delta R_{\mu/e}$ gets enhancement as $m_A$ increases, 
which is attributed to the logarithmic enhancement of $\ln(m_A^2/m_H^2)$ as discussed above. 
As a result, $m_A$ cannot exceed around 160 GeV for $|y_{\mu E}|=1.0$
while $m_A$ is allowed up to 350 GeV for $|y_{\mu E}|=0.5$ and 0.3. \footnote{There exists an upper bound on $m_A$ coming from the perturbativity of $\lambda_3$ or stability of the DM. In the IDM and its extensions, 
the latter gives the stronger bound.
In the IDM, $m_A^{\text{max}}\simeq (300-350)$ GeV modulated by the input parameters~\cite{Blinov:2015vma,Senaha:2018xek}. In the VLIDM, on the other hand, the presence of the vector-like leptons can push it upward~\cite{bar}.}
We also display the region of $v_C/T_C>1$ as the right-hand side of the vertical line~\cite{Chowdhury:2011ga,Borah:2012pu,Gil:2012ya,Cline:2013bln,AbdusSalam:2013eya,Blinov:2015vma,Cao:2017oez,Laine:2017hdk,Senaha:2018xek,Kainulainen:2019kyp}.\footnote{Note that there still exist a lot of theoretical uncertainties in ordinary perturbative treatment of EWPT~\cite{Kainulainen:2019kyp} that we adopt here based on Ref.~\cite{Senaha:2018xek}. Thus, when we interpret the $v_C/T_C=1$ line, such uncertainties are kept in mind. Nevertheless, the correlation between the enhancement of $Z\to \mu^+\mu^-$ and SFOEWPT is still robust.
}
It can been seen that SFOEWPT requires the large mass splitting of $m_H$ and $m_A$ as explained 
in the previous section, which leads to the strong correlation between the $Z$ boson decays and SFOEWPT, 
 {\it i.e.}, the stronger the SFOEWPT becomes, the more $Z\to \mu^+\mu^-$  is enhanced.
One can see that the case of $|y_{\mu E}|=1.0$ is not consistent with SFOEWPT.
It is the interesting and important question how large $\text{Im}(y_{\mu E})$ is needed for the sufficient BAU
and whether it is compatible with the results obtained here for successful baryogenesis~\cite{bar}.

Future sensitivities of $\Delta R_{\mu/e}$ are also shown by the middle and lower horizontal dashed lines,
where the former is the Giga-$Z$ at the ILC~\cite{Irles:2019xny} and the latter is Tera-$Z$ such as the CEPC~\cite{CEPCStudyGroup:2018ghi} and FCC-ee~\cite{Blondel:2018mad}. \footnote{Exactly, the sensitivities at the CEPC and FCC-ee are not necessarily the same due to different machine properties, etc. Here, we just ignore such a difference for simplicity.}
From Fig.~\ref{Rmue_MA}, it is found that the future precise measurements of the $Z$ boson can provide new and thorough tests of SFOEWPT in our benchmark scenario. This new probe can give the crosscheck of EWPT together with the conventional  approaches through the measurements of the triple Higgs boson coupling~\cite{Grojean:2004xa,Kanemura:2004ch} and gravitational waves~\cite{Huang:2017rzf}.
It should be emphasized that depending on parameter regions, the precise measurements of the $Z$ boson decays are more powerful than the conventional methods~\cite{Huang:2016odd}.
\begin{figure}[t]
\begin{center}
\includegraphics[scale=0.67]{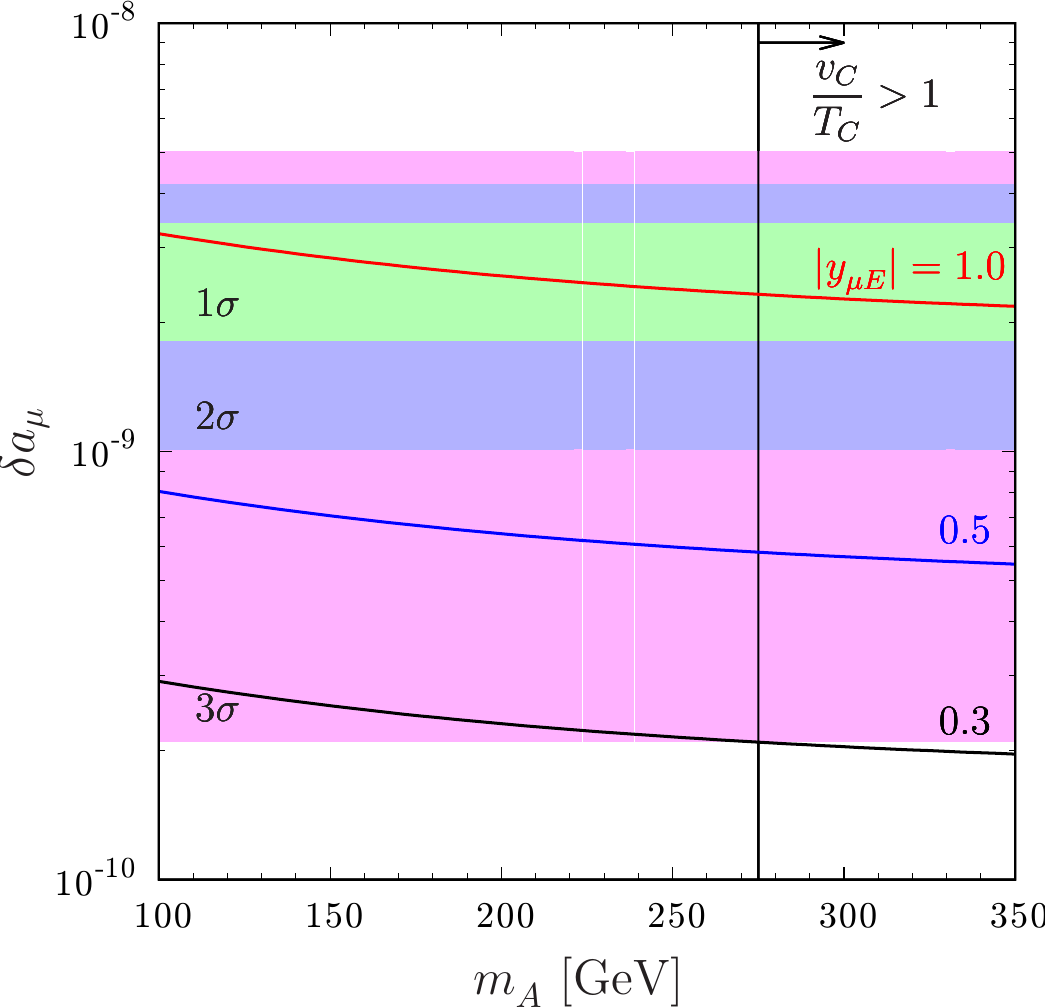}
\end{center} 
\caption{Corrections to the muon magnetic dipole moment anomaly $\delta a_\mu$ as a function of $m_A$.
The shaded regions in green, blue and magenta correspond to the $(g-2)_\mu$ regions
within 1$\sigma$, 2$\sigma$ and 3$\sigma$, respectively. The solid curves in red, blue and black correspond to the deviation $\delta a_\mu$ for $|y_{\mu E}|=1.0$, 0.5 and 0.3, respectively.}
\label{gm2_MA}
\end{figure}

Now we investigate the compatibility with $(g-2)_\mu$ in our benchmark scenario. 
In Fig.~\ref{gm2_MA}, $\delta a_\mu$ is shown as a function of $m_A$. 
The color coordinates of the solid curves are the same as those in Fig.~\ref{Rmue_MA}.
The shaded regions in green, blue and magenta correspond to the $(g-2)_\mu$ regions
within 1$\sigma$, 2$\sigma$ and 3$\sigma$, respectively.
One can see that the case of $|y_{\mu E}|=1.0$ can explain $(g-2)_\mu$ at 1$\sigma$ level.
However, as shown in Fig.~\ref{Rmue_MA}, the region of $m_A\gtrsim 160$ GeV is excluded by the measurement of $R_{\mu/e}$, thereby SFOEWPT and $(g-2)_\mu$ at the 1$\sigma$ explanation are not compatible.
We find that the explanation of $(g-2)_\mu$ is impossible within 2$\sigma$ level in the regions of SFOEWPT.

\begin{figure}[t]
\begin{center}
\includegraphics[scale=0.35]{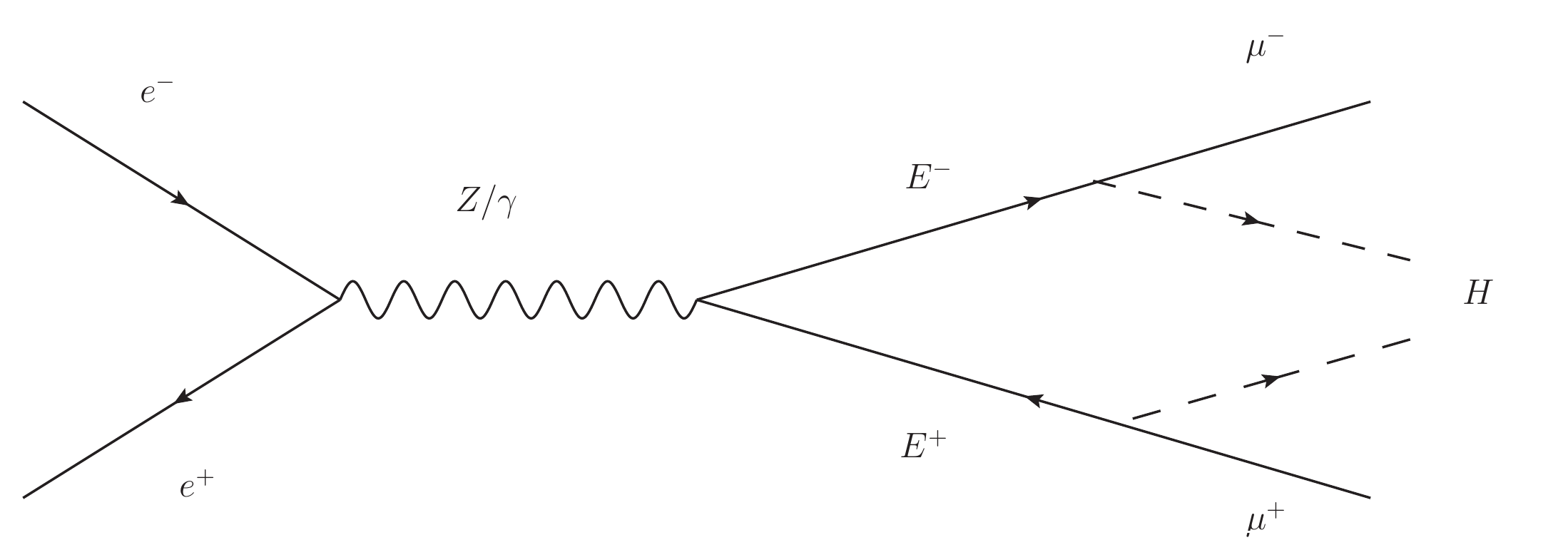}
\end{center} 
\caption{Direct search for the vector-like leptons via the process involving dimuon plus MET at the lepton colliders.}
\label{g_2_mini}
\end{figure}

Now let us consider the case in which the pair production of the vector-like leptons are kinematically allowed.
Since the vector-like leptons exclusively decay into the DM and muons, 
the most relevant process is $e^+e^-\to E^+E^- \to \mu^+\mu^- + 2H$, as shown in Fig.~\ref{g_2_mini}.
The dominant contribution to the production cross section of the vector-like leptons is 
\begin{align}
\sigma(e^+e^-\to \gamma \to  E^+E^-)=\frac{4 \pi \alpha^2}{3 s^2}  (s+2 m_E^2)\sqrt{1-\frac{4m_E^2}{s}},
\end{align}
where $\alpha$ denotes the fine structure constant at the scale of $Z$ boson mass, $s$ is the square of the center of mass energy.
Here, the masses of the electron and positrons are neglected. 

\begin{figure}[t]
\begin{center}
\includegraphics[scale=0.45]{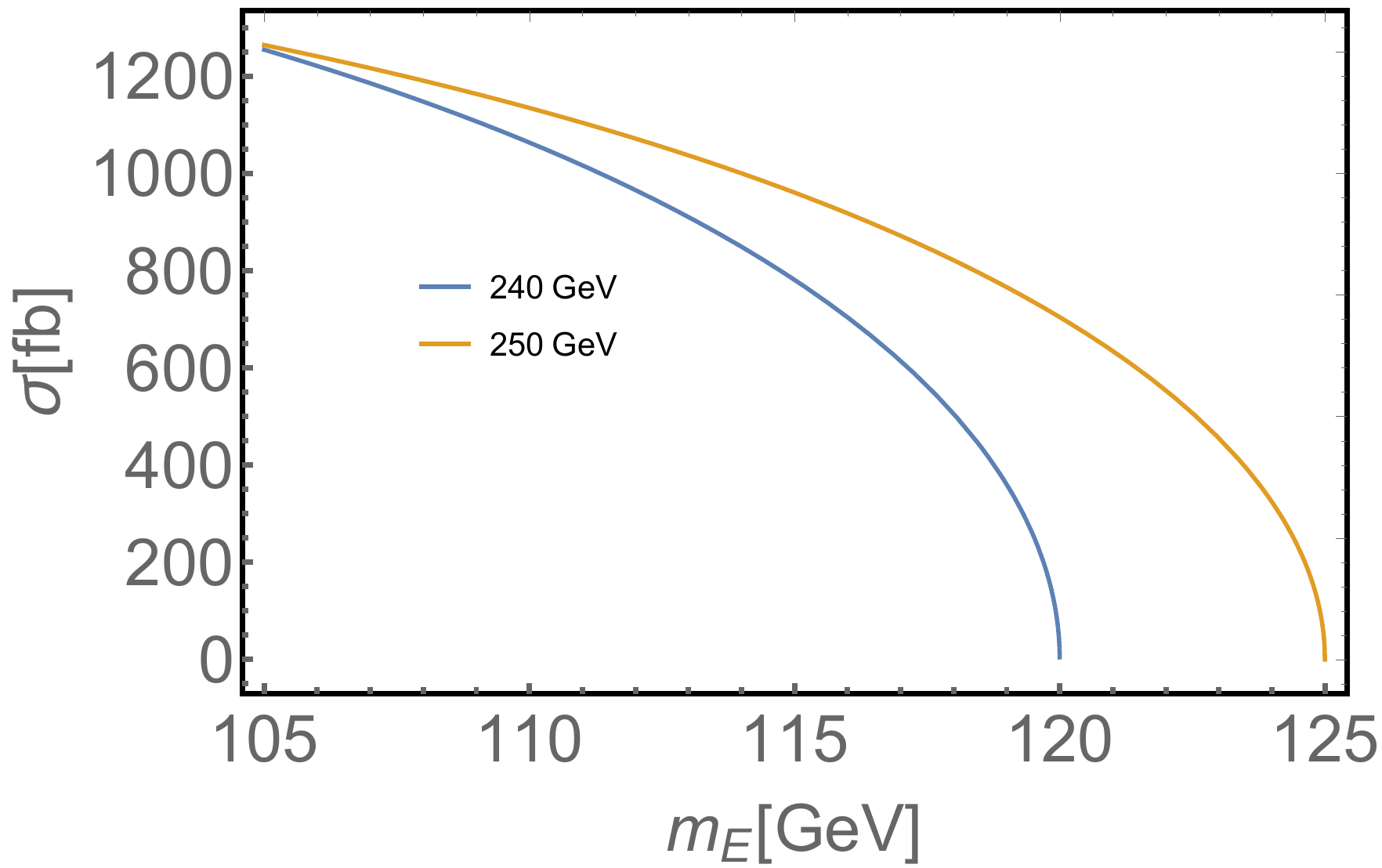}
\end{center}
\caption{The cross-section $\sigma(e^+e^-\to \gamma/Z \to  E^+E^-)$ as a function of $m_E$. We take $\sqrt{s}=240$ GeV (blue) and 250 GeV (orange), respectively.}
\label{sigeeEE}
\end{figure}

Figure~\ref{sigeeEE} shows the cross-section $\sigma(e^+e^-\to \gamma/Z \to  E^+E^-)$ as a function of $m_E$, which depends only on the collider energy and the vector-like lepton mass.
Here, we also include $\sigma(e^+e^-\to Z \to  E^+E^-)$ and the interference effects between the $Z$ boson and photon mediators. 
It is found that the cross section gets enhanced with decreasing $m_E$ and reaches about $\mathcal{O}(1)$ pb,
which is large enough to be measured at the future lepton colliders.

We can use the direct production channel to fix the vector-like lepton mass, and 
the mass splitting between the neutral scalars and the Yukawa coupling
$| y_{\mu E} |$ can be extracted from the precise measurements of the $Z$ boson decays,
and if available, together with the triple Higgs boson coupling~\cite{Grojean:2004xa,Kanemura:2004ch} and/or gravitational waves~\cite{Huang:2017rzf}.
In this way, our scenario can be fully tested.

\section{Conclusion and discussions}\label{sec:conclusion}
We have studied the possibility of SFOEWPT and its phenomenological consequences in the VLIDM. It is found that the significant mass splitting between the DM $H$ and 
the CP-odd Higgs boson $A$ is required to induce SFOEWPT. 
Such a condition inevitably leads to the enhanced $\Gamma(Z\to \mu^+\mu^-)$ owing to the logarithmic enhancement of $\ln(m_A^2/m_H^2)$. 
As a result, the couplings of the vector-like leptons to the muons have to be small in order
to evade the current experimental constraints of the $Z$ boson decays, 
which limits the size of the corrections to $(g-2)_\mu$.
Our numerical studies show that $(g-2)_\mu$ cannot be explained within 2$\sigma$ level in the region of SFOEWPT. 
In other words, EWPT would be weak first order if $(g-2)_\mu$ is confirmed in this model.

We also showed that the precise measurement of $Z\to \mu^+\mu^-$ at future lepton colliders, 
such as the ILC Giga-$Z$ and CEPC/FCC-ee Tera-$Z$ as well as the direct search of the vector-like leptons
via the process $e^+e^-\to \gamma/Z\to E^+E^-$ can provide new exquisite probes of SFOEWPT.

It should be emphasized that the deep correlation between SFOEWPT and the enhancement of the $Z$ boson decays 
would be generic as long as SFOEWPT requires the large mass splitting of the scalar mass spectrum in the same multiplet, which opens a novel and promising avenue to probe thermal history of EWPT in the early Universe and EWBG in addition to the well-studied approaches using the triple Higgs coupling and gravitational waves.

As a by-product, our study clarified that the future lepton colliders, especially the $Z$ factories, can provide
new alternative approach to explore the DM blind spots, where the DM-Higgs coupling $\lambda_L$ is too small
to be detected by the DM direct detection.

Lastly, we make a remark about other DM scenarios such as the heavy scalar DM ($m_H\gtrsim 500$ GeV)~\cite{Banerjee:2019luv} and the right-handed neutrino DM~\cite{Vicente:2014wga}.
In the former case, SFOEWPT induced by the thermal loop effects would not be realized since the conditions of $\mu_2^2\ll \lambda_3v^2/2$ 
cannot be satisfied. 
In the latter case, on the other hand, $\lambda_5$ has to be much smaller than the unity to be consistent with neutrino and DM physics, which implies that $m_{H^\pm}=m_A\simeq m_H$.
It is still possible to have SFOEWPT as long as $\mu_2^2\ll \lambda_3v^2/2$.
However, its correlation with the $Z$ decays would be lost due to lack of the large mass splitting between $H$ and $A$. 

\begin{acknowledgments}
This work was supported by IBS under the project code, IBS-R018-D1.
\end{acknowledgments}

\appendix
\section{Loop functions}\label{LF}
The one-loop functions appearing in $\Delta g_{Z\bar{\ell}\ell}^{L}$ and $\Delta g_{Z\bar{\nu}\nu}^{L}$ are defined as 
\begin{align}
F_2(m_{E}, m_\phi) & = \int_x~x\ln\Big[(1-x)m_{E}^2+xm_\phi^2 \Big], \\
F_3(m_{E}, m_\phi) & = \int_{xy}~
\bigg[
\frac{xym_Z^2+m_{E}^2}{\Delta_3}-1-\ln\Delta_3
\bigg],\\
\tilde{F}_3(m_{E},m_H,m_A) &= \int_{xy}~\ln\tilde{\Delta}_3,
\end{align}
where $\int_x=\int_0^1dx$, $\int_{xy}=\int_0^1dx \int_0^{1-x}dy$ and 
\begin{align}
\Delta_3 &= -xy m_Z^2+(x+y)m_{E}^2+m_\phi^2(1-x-y), \\
\tilde{\Delta}_3 &= -xym_Z^2+xm_H^2+ym_A^2+m_{E}^2(1-x-y).
\end{align}
Incidentally, for $m_H=m_A=m_\phi$, our loop functions are reduced to those in Ref.~\cite{Calibbi:2018rzv}:
\begin{align}
I_a(m_{E},m_\phi) &= F_3(m_{E},m_\phi)+\tilde{F}_3(m_{E},m_\phi), \\
I_c(m_{E},m_\phi) &= F_2(m_{E},m_\phi)-F_3(m_{E},m_\phi).
\end{align}


\end{document}